# Bridge2AI Recommendations for AI-Ready Genomic Data

Matthew Cannon[1*], Wesley Goar[1*], In-Hee Lee[3], James Stevenson[1], Amy Heiser[5], Nathan Sheffield[6], James Eddy[7], Monica Munoz-Torres[8], Sek Won Kong[3,4], Alex H. Wagner[1,2,†]

[1] Institute for Genomic Medicine (IGM), Nationwide Children's Hospital, Columbus, OH

[2] Department of Pediatrics, The Ohio State University College of Medicine, Columbus, OH

[3] Computational Health Informatics Program, Boston Children's Hospital, Boston, MA

[4] Department of Pediatrics, Harvard Medical School, Boston, MA

[5] Sage Bionetworks, Seattle, WA

[6] University of Virginia, Charlottesville, VA

[7] Avantiqor, Washington, DC

[8] Department of Biomedical Informatics, University of Colorado Anschutz Medical Campus, Aurora, CO

* authors contributed equally

† Correspondence:

Alex Wagner, PhD, Nationwide Children's Hospital, Room WB3155, Research Building 3, 575 Children's Crossroad, Columbus, OH 43210. Phone: (614) 355-1645, Email: Alex.Wagner@nationwidechildrens.org

**ABSTRACT**

Rapid advancements in technology have led to an increased use of artificial intelligence (AI) technologies in medicine and bioinformatics research. In anticipation of this, the National Institutes of Health (NIH) assembled the Bridge to Artificial Intelligence (Bridge2AI) consortium to coordinate development of "AI-ready" datasets that can be leveraged by AI models to address grand challenges in human health and disease. The widespread availability of genome sequencing technologies for biomedical research presents a key data type for informing AI models, necessitating that genomics data sets are "AI-ready". To this end, the Genomic Information Standards Team (GIST) of the Bridge2AI Standards Working Group has documented a set of recommendations for maintaining AI-ready genomics datasets. In this report, we describe recommendations for the collection, storage, identification, and proper use of genomics datasets to enable them to be considered "AI-ready" and thus drive new insights in medicine through AI and machine learning applications.

# INTRODUCTION

*Motivation*

State-of-the-art artificial intelligence (AI) technologies are now being applied to biology and bioinformatics research problems across the world to great effect. Despite their widespread adoption and applications for large biological (and clinical) datasets, the definition of what an AI-ready dataset looks like has only recently been discussed. Led by the U.S. National Institutes of Health (NIH) Bridge to Artificial Intelligence (Bridge2AI) consortium, AI-ready datasets are those that are FAIR, fully reliable, robustly defined, and computationally accessible[1]. Using these assessments as a guide, Bridge2AI data generation projects such as "Cell Maps for Artificial Intelligence" (CM4AI) have created robust, reusable, and explainable biomedical datasets that can be used with AI and machine learning (ML) applications to improve our understanding of human health and our ability to treat human disease[2]. Given the power of AI technologies for generating new insights in medicine, our ability to prepare datasets and provide provenance for these applications will be essential for driving new discoveries in existing medical fields.

Advances in sequencing technologies and bioinformatics have led to an exponential increase in the amount of genomic sequencing data being generated for biomedical studies[3,4]. Accompanying this growth is a commensurate opportunity for enabling these data (and associated downstream data such as gene expression quantification and sequence variant calls) for use in AI applications. However, the diversity of sequencing methods varies wildly, creating barriers to automated reuse of these data to drive meaningful breakthroughs and insights via AI/ML applications. Explainability and reproducibility at all stages of data collection and model development is therefore essential to compel trust in discoveries generated through AI/ML affiliated methodologies[5,6].

To guide development of the Bridge2AI data generation projects and broader community efforts to construct AI-ready genomic datasets, the Bridge2AI Genomic Information Standards Team (GIST) has developed the following summary of data and metadata recommendations for genomic sequencing experiments to support AI-driven biomedical discovery. These recommendations build upon the principles of AI-readiness for biomedical data[1], and broadly describe the sufficient metadata necessary for robust and explainable datasets that are suitable for use in downstream AI applications. In this work, we cover the collection, storage, identification, and use of genomic sequencing data to meet these criteria. The remit of these recommendations is genomic DNA sequencing data (i.e., "sequencing reads" and downstreamed artifacts), which should be preserved for reanalysis. While consent and clinical phenotype metadata should be included, we will not discuss these to keep our recommendations modular and focused on genomic data and its pre- and post-processing.

**MAIN TEXT**

Central to AI-readiness is the need for the data to contain the necessary metadata for automated and manual interpretation, evaluation, and validation. The metadata we describe below is what must or should be included to control for covariates during the processing of genomic sequence data. Definitions for genomics metadata terminology and file formats throughout this report have been assembled from established ontologies and are provided in **Supplementary Table 1**.[7–9] A more detailed list of possible metadata to ensure findable, accessible, interoperable, and reusable (FAIR) genomes can be found at FAIR Genomes, EDAM, MIxS GSC, Sequence Ontology, EFO, and NCIT, and should be used when possible[9–13]. All datasets must include a metadata file describing 1) what each file in the dataset contains, 2) the date the dataset was generated, 3) data processing methods used, 4) site where the file was processed, and 5) an email address for the study's contact individual. The description of file contents should describe each data field with corresponding data types using community-accepted metadata standards while ensuring privacy. In the following sections, we provide recommendations for relevant metadata fields to include, as well as references to community-accepted standards for each data type and example values where applicable.

*Description of supporting metadata for sample origin and preparation*

Characteristics of the originating sample used for genomic sequencing that describe the source, sample quality and provenance (summarized in **Table 1**) are necessary for informing downstream analyses. Sample preservation and storage conditions must be included and should describe in detail how the genomic sample was stored and maintained, e.g., if samples were fresh, flash frozen, or stored in formalin-fixed paraffin-embedded blocks. These metadata inform downstream AI and human agents on processing limitations and potential biases inherent in the source sample; for example, formalin-fixed tissue often undergoes more DNA degradation than other storage methods, such as flash frozen samples[14], adversely affecting read quality. We also recommend the inclusion of the dates and locations of sample collection and sample preparation, the sampling protocol, biospecimen type (e.g., tissue or cell culture), clinical diagnosis, and pathological state (e.g., normal or tumor), all of which should be included within this subset of metadata (**Table 1**). If the sample is derived from a human clinical specimen, the metadata should include sex assigned at birth, genetic ancestry, and clinical phenotypes.

*Description of sequencing preparation and process*

Structured metadata capturing sequencing library preparation and instrumentation support the explainability and readiness of genomic data for AI applications (**Table 2**). This includes capturing the library preparation and sequencing process (e.g., exome capture methods and PCR amplification), as well as the region coverage. If targeted panel or exome sequencing was performed, a browser extensible data (BED) file[15] should be provided that supplies the profiled regions/locations to control for off-target or low sequence coverage results. Enrichment kits used for targeted sequencing experiments operate by different methods that should be described. Data producers should note what barcodes, unique molecular identifiers, and other synthetic sequences were used in this experiment, to inform the reanalysis and reuse of these data in downstream analysis pipelines. Metadata must also include information on the sequencing platform and instrument model, sequence pool and run identifiers, and the date and location each sample was sequenced. If applicable, samples sequenced across runs on multiple different machines should include these metadata for each applicable run contributing to the generated data.

*Description of specifications for full genomic sequencing procedure*

Sequencing data analysis pipelines are highly variable and essential to the interpretation of results generated from a sequencing experiment. Data analysis workflows often include bespoke data transformation and analysis methods that are definitional to the downstream variant call results. To control for some of these differences, we recommend including metadata specific to the processing and analysis of sequencing results (**Table 3**). Understanding the inherent purpose or intention behind an experiment is required to understand sequencing data. Data producers must include the analyses performed (e.g., phased genotyping, chromatin profiling, clonal variant calling) and the bioinformatics workflow used to produce the provided results files.

Pipeline information should include the software used throughout the pipeline including the computational environment and version of the software (lockfile) with parameters used (including random seeds), and the specific algorithms (including deviations or modifications) used. If the data has been aligned to a reference genome (e.g., GRCh38), this must be included along with the reference version (e.g,. GCF_000001405.40) and pointer to the source of the reference genome. Metadata about reference sequences should be stored alongside downstream alignments and variant calls for genomic data. Standard sequence digest approaches should be used to create identifiers for sequences, and digests should be stored in databases and file headers when used to produce file contents. We recommend that URL-safe sha512t24u[16,17] reference sequence digests be used for this purpose.

Specifications for hardware (e.g., GPU or FPGA) and the analysis environment used during the pipeline are recommended to be included to control for known/reported issues with these and the software being used for processing. Within the resulting data set, a descriptive file must be included that describes what each file in the dataset should contain (including format and descriptions), and the date and location the file was processed. For example, if that dataset contains an annotated gVCF, this file should clearly explain what each column header means and the intended format and content of data in that column. We also recommend that consent and data use permissions be included to ensure data is not legally being misused; however, those data are ubiquitous to datasets and are outside the scope of these recommendations.

*Capturing adequate quality control data*

Data that has been aligned should include base call and read mapping quality, paired read data (as applicable) and depth of coverage scores (**Table 4**). The ideal depth of coverage scores are calculated post-deduplication to accurately reflect the usable reads for variant calling. Scores should include mean coverage, and if the experiment is a targeted capture, the quality control data should also capture the percent of target bases with suitable coverage (e.g., 30x depth), as this is a key metric for assessing the quality of the capture kit and sequencing run. Other quality metrics that should be provided include 1) error rate summary, 2) sequenced GC content, 3) breadth of coverage (for alignment data), and 4) chromosomal ploidy, if known (including sex chromosomes).

*Data storage formats*

Appropriate file formats to store genome sequences include FASTQ (reads only), BAM (reads with alignments), and CRAM (compressed reads with alignments). Read data only needs to be preserved in one of these formats. While BAM and CRAM files allow for storing alignments without accompanying read sequence or base quality scores, we recommend that sequence read and alignment data always be stored together in CRAM files (**Table 5**).

Variant call data should include variant quality scores, the total number of reads covering the variant, the number of reads supporting a reference allele call, and the number of reads supporting an alternate allele call (**Table 6**). As highlighted in the NHGRI FAIR Data Workshop, for semantically precise representation of variant data for use by AI systems, we recommend the use of data standards compatible with the GA4GH Variation Representation Specification (VRS)[17]. The VRS standard provides a semantically-rich schema that works as a *lingua franca* for representing all forms of genomic variation, an information model designed for cross-resource genomic data integration, and a standard method for generating globally unique computational identifiers.

For storing small variant (e.g., SNVs and indels) data from large-scale sequencing studies, we recommend these data are stored as gVCF files, based on the GA4GH Variant Call Format (VCF) (at least version 4.2), with version 4.3 preferred. Users may also include other downstream formats, such as GFF3 for annotated data. For pipelines using the GA4GH VCF file format, these files can be readily made VRS-compatible using the VRS-Python software stack.

*Autism genotype-phenotype associations as an example of the importance of metadata capture*

Proper capture and documentation of metadata is essential for enabling genomics and precision medicine through AI/ML operations. Without the proper context and procedural details, an AI model trained using ill-described data runs the risk of generating insights linked to experimental artifacts rather than true biological effects. To illustrate this, we have provided a sample use case in the observation of Autism genotype-phenotype associations (**Supplementary Data**). In brief, we demonstrate how the absence of key metadata from WGS samples used to train an AI model resulted in preliminary findings that were more closely associated with sample source (saliva vs. blood) than any real biological associations. By demonstrating these risks, we hope to further persuade the proper capture and standardization of metadata generated from genomics approaches to enable robust AI/ML applications.

*Conclusion*

The emergence of widespread AI technologies has created opportunities for the application of genomics data in scalable, machine-driven biomedical discovery. Alongside this potential, however, also comes a new responsibility to ensure that genomics datasets are AI-ready; that is, data must be explainable, reusable, and computationally available for precise usage. The recommendations and standards for metadata discussed in this report represent reliable consensus guidelines that, when followed, will improve the AI/ML-ready status of genomic sequencing datasets. The primary limitation of these recommendations are their limited scope, which is focused specifically on small variant detection, and therefore does not fully capture all the variability present in genomics research. However, we believe that many of these recommendations are generalizable beyond small variant studies, and encourage following the guidance of the GA4GH for additional recommendations in preparing AI-ready genomics data.

# FIGURES AND TABLES

| Requirement Level | Name | Existing Standards | Sample Derived From |
|---|---|---|---|
| Must | Storage Conditions | 1,3,5,6 | All |
| Must | Date of Sample Collection | 1,3,5,6 | All |
| Must | Location of Sample Collection | 1,6 | All |
| Must | Date of Sample Preparation | 1 | All |
| Must | Location of Sample Preparation | 1,6 | All |
| Must | Sampling Protocol | 1,5,6 | All |
| Must | Sample ID (De-identified) | 1,2,3,6 | All |
| Must | Biospecimen Type | 1,6 | All |
| Must | Clinical Diagnosis | 1,6 | All |
| Must | Pathological State | 1,6 | All |
| Should | Sex assigned at Birth | 1,3,7 | Human |
| Should | Genetic Ancestry | 1,3,5,6 | Human |
| Should | Phenotypes | 1,6 | Human |
| Should | Anatomical Source of the Biospecimen | 1,3,6 | All |

Standards Legend: 1. FAIR Genomes  2. EDAM  3. MIxS GSC  4. Sequence Ontology  5. EFO  6. NCIT  7. GSSO

**Table 1. Metadata Recommendations to Describe Sample Origin and Preparation**

| Requirement Level | Name | Existing Standards | Sample Derived From |
|---|---|---|---|
| Must | Library Preparation | 1,3,5,6 | All |
| Must / If Applicable | Targeted Coverage Regions (BED file) | 2,6 | All |
| Should / If Applicable | Enrichment Kit Used | 1,3,6 | All |
| Must | Were Unique Molecular Identifiers Used? | 1,3,5,6 | All |
| Must | Sequencing Platform | 1,3,6 | All |
| Must | Sequencing Instrument Model | 1,3,6 | All |
| Must | Sequencing Method | 1,3,6 | All |
| Must | Sample Sequencing Date | 1,5,6 | All |
| Must | Sample Sequencing Location | 1,2,3,6 | All |
| Should | Sequencer ID |  | All |

**Table 2. Metadata Recommendations to Describe Sequencing Preparation and Process**

Standards Legend: 1. FAIR Genomes  2. EDAM  3. MIxS GSC  4. Sequence Ontology  5. EFO  6. NCIT

| Requirement Level | Name | Existing Standards | Sample Derived From |
|---|---|---|---|
| Must | Analyses Performed | 1,2 | All |
| Must | Software Used for Bioinformatics Analyses | 1,2,5,6 | All |
| Must | Version of Software used for Bioinformatics Analyses | 2,6 | All |
| Must | Parameters used for each step of the Bioinformatics Analyses | 1,2,6 | All |
| Must | Specific Algorithms used during the Bioinformatics Analyses | 1,2,6 | All |
| Must | Deviations or Modifications to Algorithms used | 1,6 | All |
| Must | Reference Genome used for Alignment (include pointer to source file) | 1,2,3,6 | All |
| Must | Sequence Identifiers | 6,7 | All |
| Should | Specifications for Hardware | 5,6 | All |
| Should | Analysis Environment used | 6 | All |
| Must | Date processed (every file) | Most have a general date | All |
| Must | Location processed (every file) | Most have a general location | All |

**Table 3. Metadata recommendations to describe genomic sequencing processing and procedure**

Standards Legend: 1. FAIR_Genomes 2. EDAM 3. MIxS_GSC 4. Sequence Ontology 5. EFO 6. NCIT 7. GA4GH

| Requirement Level | Name | Example Value | Sample Derived From |
|---|---|---|---|
| Must | Read Mapping Quality | 10 | All |
| Must | Depth of Coverage Scores | 45 | All |
| Should / If Applicable | Mean Coverage | 100 | All |
| Should / If Applicable | Percent of Target Bases with Suitable Coverage | 75% | All |
| Should | Error Summary Metrics | '.error_by_base_quality' file | All |
| Should | GC Content | 25% | All |
| Should | Breadth of Coverage (alignment data) | 95% | All |
| Should | Genotypic Sex | XX Genotype | Human |

**Table 4. Recommendations for Quality Control Metrics**

| Requirement Level | Data Storage Formats | Contains | Sample Derived From |
|---|---|---|---|
| Must | CRAM/BAM/SAM | Alignment data | All |
| Must | gVCF (min version 4.2) | Genomic variant calls | All |
| Must | FASTA | Nucleotide sequences | All |
| Must | FASTQ | Reads with quality scores | All |
| Must | GFF | Genomic feature annotations | All |

**Table 5. Data storage formats**

| Requirement Level | Name | Example Value | Sample Derived From |
|---|---|---|---|
| Must | Variant Quality Score | PASS/FAIL (can contain other specific codes) | All |
| Must | Total Reads Covering the Variant | 150 | All |
| Must | Number of Reads Supporting a Reference Allele Call | 75 | All |
| Must | Number of Reads Supporting an Alternate Allele Call | 75 | All |
| Must | VRS Variant Object Data | <VRS Location / VRS State> | All |
| Should | VRS Identifiers | ga4gh:VA.EgHPXXhULTwoP4-ACfs-YCXaeUQJBjH_ | All |
| May | Genotype Quality Score | 89 | All |

**Table 6. Metadata Recommendations to Support Variant Call Data**

*Use Case: The Essential Role of Metadata in Refining Genotype-Phenotype Associations for Autism Sequencing Projects*

Objective

To illustrate the critical value of metadata in discovering the genotype-phenotype association accurately, demonstrated in the context of a large-scale autism sequencing project undertaken by genome centers B1 and B2.

Background

The project embarked on whole-genome sequencing (WGS) to investigate the genetic intricacies of autism. Despite advanced sequencing and analytical methods, initial analyses by B1 and B2 faced challenges in identifying genotype-phenotype associations accurately.

Initial Approach and Challenges

- Due to the quality of the collected sample and concern of contamination during sample collection, WGS conducted in both centers primarily used DNA extracted from blood, with a subgroup using DNA from saliva samples. However, the proportion of saliva samples was different between B1 and B2
- Each center initiated WGS with slight variations in PCR chemistry: B1 utilized 2-channel while B2 employed a mix of 2-channel and 4-channel chemistry sequencers.
- Different software pipelines were used for alignment and variant calling across centers, introducing variability in data processing: BWA-MEM and GATK at B1, and DRAGMAP and DRAGEN at B2.
- The phenotypic measures comprised 27 variables, both categorical and continuous, for neuro-cognitive development.

An initial AI model was developed, pre-trained on population-scale genomic variants and phenotypes, and then fine-tuned with autism WGS data from both centers. However, the omission of metadata related to sample sources (saliva vs. blood), along with subtle phenotypic score differences caused by discrepancies in sequencing and data processing, resulted in preliminary findings that were linked to technical artifacts rather than to genuine biological associations.

Metadata Integration

Recognizing the need to overcome these challenges, a second AI model was conceptualized with a comprehensive metadata approach:

1. **Pre-training** involved both sample-level and variant-level metadata, including details about the sequencing chemistry and procedures unique to centers B1 and B2.
2. **Fine-tuning** was enhanced with WGS metadata, encompassing sequencing depth, alignment, and variant calling nuances, alongside the 27 neuro-cognitive development measures.

Outcomes

The refined model significantly outperformed its predecessor. Associations previously misconstrued as biologically relevant, when they were actually artifacts of the variances in sequencing methods and sample sources, were filtered out successfully. The refined model identified several new promising genotype-phenotype associations, which are anticipated to impact patient care. It highlighted how the associations identified in the first model were predominantly linked to technical variables such as DELIN sequences, homopolymers, and center-specific SNVs rather than true genotype-phenotype correlations.

## Impact

- **Precision in Discovery:** The integration of detailed metadata directly addressed the initial model's limitations, leading to a significant improvement in identifying authentic and novel genetic associations.
- **Insights into Methodological Variability:** The use case underscored the nuanced effects of methodological differences between sequencing centers on research outcomes, advocating for a standardized approach to data collection and analysis.
- **Guidance for Future Genomic Research:** This project demonstrates the critical role of metadata in genomic studies, offering a blueprint for future projects to ensure accuracy and reliability in their findings.

## Conclusion

The autism sequencing project's journey from initial challenges to breakthrough discoveries, facilitated by the strategic inclusion of metadata, illustrates the transformative power of detailed data contextualization in genomic research. This use case not only champions the meticulous inclusion of metadata but also sets a precedent for future endeavors in the field, aiming to make genomic studies as precise and informative as possible.

As clearly demonstrated by the included use case, to enable precision medicine through AI/ML operations, genomic sequencing data generators need to include key metadata and experimental data. The inclusion of such data allows for the control of covariates and makes it possible to tease out real results vs artifacts when performing AI/ML operations.

| Term | Ontology/Reference |
|---|---|
| sampling protocol | https://www.ebi.ac.uk/ols4/ontologies/efo/classes?short_form=EFO_0005518 |
| biospecimen type | https://ontobee.org/ontology/NCIT?iri=http://purl.obolibrary.org/obo/NCIT_C70713 |
| key clinical diagnosis | https://ontobee.org/ontology/NCIT?iri=http://purl.obolibrary.org/obo/NCIT_C15607 |
| pathological state | https://github.com/fairgenomes/fairgenomes-semantic-model/blob/main/lookups/PathologicalState.txt |
| sex assigned at birth | https://ontobee.org/ontology/GSSO?iri=http://purl.obolibrary.org/obo/GSSO_009418 |
| genetic ancestry | https://ontobee.org/ontology/NCIT?iri=http://purl.obolibrary.org/obo/NCIT_C176763 |
| phenotypes | https://ontobee.org/ontology/NCIT?iri=http://purl.obolibrary.org/obo/NCIT_C16977 |
| library preparation | https://github.com/fairgenomes/fairgenomes-semantic-model/blob/main/lookups/NGSKits.txt |
| unique molecular identifiers | https://www.ebi.ac.uk/ols4/ontologies/efo/classes?short_form=EFO_0010199 |
| sequencing platform | https://github.com/fairgenomes/fairgenomes-semantic-model/blob/main/lookups/SequencingPlatform.txt |
| sequencing instrument model | https://github.com/fairgenomes/fairgenomes-semantic-model/blob/main/lookups/SequencingInstrumentModels.txt |
| sequencing method | https://github.com/fairgenomes/fairgenomes-semantic-model/blob/main/lookups/SequencingMethods.txt |
| analyses performed | https://github.com/fairgenomes/fairgenomes-semantic-model/blob/main/lookups/AnalysesPerformed.txt |
| reference genome | https://github.com/fairgenomes/fairgenomes-semantic-model/blob/main/lookups/GenomeAccessions.txt |
| genotypic sex | https://github.com/fairgenomes/fairgenomes-semantic-model/blob/main/lookups/GenotypicSex.txt |
| FASTQ format | https://maq.sourceforge.net/fastq.shtml |
| BAM format | https://samtools.github.io/hts-specs/SAMv1.pdf |
| CRAM format | https://samtools.github.io/hts-specs/CRAMv3.pdf |

**Supplementary Table 1. Supporting definitions and ontology references for genomics metadata terminology.**